\documentclass{osa-article}

\journal{osac}

\articletype{Research Article}

\begin{document}

\title{Towards an integrated platform for characterizing laser-driven, isochorically-heated plasmas with 1~$\mu$m spatial resolution}

\author{C. H. Allen\authormark{1,*}, M. Oliver\authormark{2}, L. Divol\authormark{3}, O. L. Landen\authormark{3}, Y. Ping\authormark{3}, M. Sch{\"o}lmerich\authormark{3}, R. Wallace\authormark{3}, R. Earley\authormark{4}, W. Theobald\authormark{4}, T. G. White\authormark{1}, and T. D{\"o}ppner\authormark{3}}

\address{\authormark{1}Physics Department, University of Nevada Reno, 1664 N Virginia St, Reno, NV 89557, USA\\
\authormark{2}Rutherford Appleton Laboratory, Didcot OX11 0DE, UK\\
\authormark{3}Lawrence Livermore National Laboratory, L-493, 70000 East Avenue, Livermore, CA 94550, USA\\
\authormark{4}Laboratory for Laser Energetics, 250 E River Rd, Rochester, NY 14623, USA}

\email{\authormark{*}challen@unr.edu} 



\begin{abstract}
Warm dense matter is a region of phase space that is of high interest to multiple scientific communities ranging from astrophysics to inertial confinement fusion. Further understanding of the conditions and properties of this complex state of matter necessitates experimental benchmarking of the current theoretical models. We discuss the development of an x-ray radiography platform designed to measure warm dense matter transport properties at large laser facilities such as the OMEGA Laser Facility. Our platform, Fresnel Diffractive Radiography, allows for high spatial resolution imaging of isochorically heated targets, resulting in notable diffractive effects at sharp density gradients that are influenced by transport properties such as thermal conductivity. We discuss initial results, highlighting the capabilities of the platform in measuring diffractive features with micron-level spatial resolution.
\end{abstract}

\section{Introduction}
Warm dense matter is a highly complex region of plasma phase space that is relevant for a wide range of fields from astrophysics to fusion energy. The warm dense matter (WDM) state is found naturally in several astrophysical environments, e.g., planetary interiors and white dwarfs\cite{Guillot1999,Paquette1986}{}. It is also of particular interest for inertial confinement fusion (ICF) efforts, as the fuel in ICF implosions to ignition passes through this phase space\cite{Atzeni2004,Hurricane2014,Hu2016}. Given this ubiquity, the WDM regime remains an active area of research, with both experimental and theoretical interests and challenges.

Primarily, there is a considerable body of theoretical and computational work for matter at WDM conditions\cite{Grabowski2020}{}. However, modeling matter in this parameter regime is particularly challenging as one often encounters systems with strong ion-ion correlations and electrons that exhibit distinct quantum behavior\cite{Graziani2014}{}. Theoretical descriptions typically have their roots in plasma or condensed matter limits, which start to break down at elevated densities or temperatures. Modern simulations, such as quantum molecular dynamics, calculate from first-principles but are computationally expensive, and have limits on the number of particles within the simulation as well as the overall simulation time \cite{Hu2016}. This can sometimes lead to an inadequate sampling of phase-space and an unphysical sensitivity to initial conditions. Further experimental verification of various properties can help benchmark these theoretical models and ensure accurate modelling.

Direct observation of WDM remains difficult, owing to the extreme conditions it exists at. No natural sources of WDM exist on Earth's surface, and thus it is limited to laboratory creation; large-scale laser facilities such as the OMEGA Laser Facility or the National Ignition Facility can create WDM from laser-produced plasmas transiently. Furthermore, WDM is opaque to visible light and therefore requires methods such as x-ray radiography or spectroscopy to probe the sample. These characteristics make direct experimentation on WDM challenging, and specifically improving our understanding of the transport properties (e.g., thermal conductivity, diffusivity, and viscosity) at these conditions is essential. Previous experiments on thermal conductivity in WDM have attempted to measure the heat flow across a sample directly utilizing techniques based on plasma emission\cite{McKelvey2017,Ping2019,Sugimoto2017}{}. While several of these experiments have been successful, they are limited to a handful of materials (Al, C, W), and the uncertainties in the measurements remain large. This is, in part, due to the complexity of analyzing the spectroscopic results where atomic models, equation-of-states, and surface effects all play a role.

 Here we describe the design of a novel experimental platform to measure thermal conductivity in WDM suitable for a variety of materials and plasma temperatures. 
 The experiment measures the evolution of the density profile at isochorically-heated WDM interfaces using phase contrast x-ray radiography. Isochoric heating is achieved by irradiation with strong x-ray sources that generally lead to temperature and hence pressure gradients at the interface. After pressure equilibration between the two sample materials, typically within a few nanoseconds, the density profile across the interface is dominated by thermal conduction. We aim to measure this profile with 1~$\mu$m spatial resolution using Fresnel Diffractive Radiography (FDR), allowing the materials' thermal conductivity to be extracted. This paper describes the technical details for both the isochoric heating geometry required to generate appropriate WDM conditions, and the alignment precision in our  radiographic imaging setup. We discuss proof of principle measurements of our target interfaces and compare against simulated profiles, highlighting the spatial resolution and capabilities of our platform.

\section{Isochoric Heating Platform}

Measuring the transport properties at a material interface requires that the interface evolves symmetrically. A non-uniform evolution can lead to instabilities or shockwaves that can quickly destabilize the interface or otherwise prevent measurement of the transport properties in question. Additionally, any non-uniformity in the plasma parameters along the imaging axis, such as temperature gradients resulting from spot heating, will increase the uncertainty in the measurement. These constraints make a buried, concentric wire system an ideal target geometry. If the target is irradiated near-uniformally, the evolution of the interfaces will be concentrically symmetric. This wire system has an additional benefit of allowing us to image two interfaces as opposed to just one, i.e. the left and right interfaces of the inner wire easily with an appropriate imaging magnification.

\begin{figure}[]
\centering
  \includegraphics[width=1\linewidth]{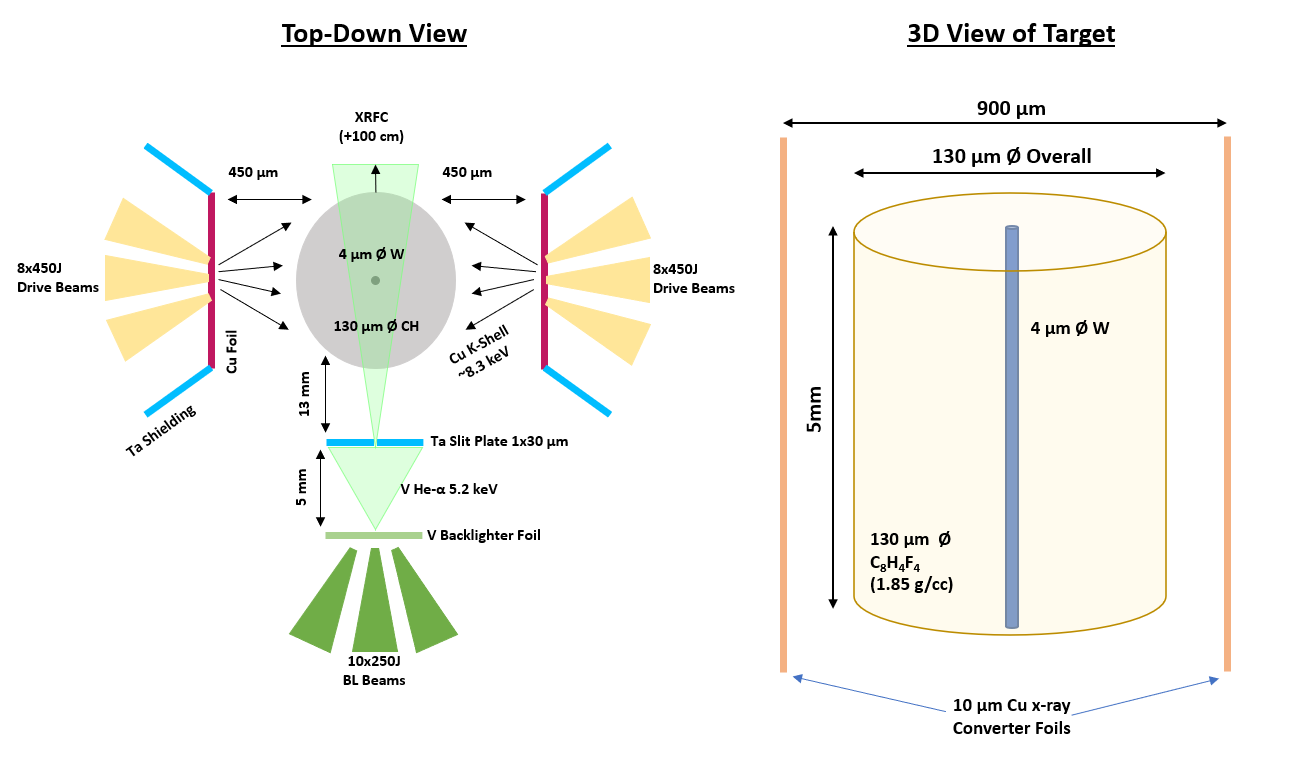}
  \caption{\small{{\bf Schematic of the experimental configuration.} A 4~$\mu$m W wire inside 130~$\mu$m diameter $\text{C}_{\text{8}}\text{H}_{\text{4}}\text{F}_{\text{4}}$ plastic is isochorically heated to warm dense matter conditions by x-rays generated through irradiation of thin Cu foils placed on either side of the target. The W wire's high opacity means it acts as a buried heater, absorbing significantly more energy than the surrounding cladding. After pressure equilibration between the samples, the subsequent evolution of the interface is primarily driven through thermal conductivity. We probe the density profile across the interface with Fresnel Diffractive Radiography using V He-$\alpha$ x-rays collimated by a 1~$\mu$m slit.}}
  \label{fig:setup}
\end{figure}

A diagram of our isochoric heating platform as fielded at the OMEGA Laser Facility at LLE Rochester \cite{OMEGAFacility} is shown in Fig. \ref{fig:setup}. The target consists of a buried 4~$\mu$m tungsten (W) wire coated with 65~$\mu$m radius $\text{C}_{\text{8}}\text{H}_{\text{4}}\text{F}_{\text{4}}$ plastic ("CH") via physical vapor deposition. Two 10~$\mu$m thick copper (Cu) foils positioned 450~$\mu$m on either side of the cladding are irradiated by sixteen 450~J, 351~nm, 1~ns square laser pulses, creating a source of Cu He-$\alpha$ emission (8.3~keV) that isochorically heats the sample. 
The W wire is chosen for its high atomic mass and opacity, while the coating is chosen for its low opacity and scientific interest. This difference in both opacity and density creates a large temperature differential across the interface, creating perfect conditions to observe the impact of thermal conductivity.

\begin{figure}[]
\centering
  \includegraphics[width=1\linewidth]{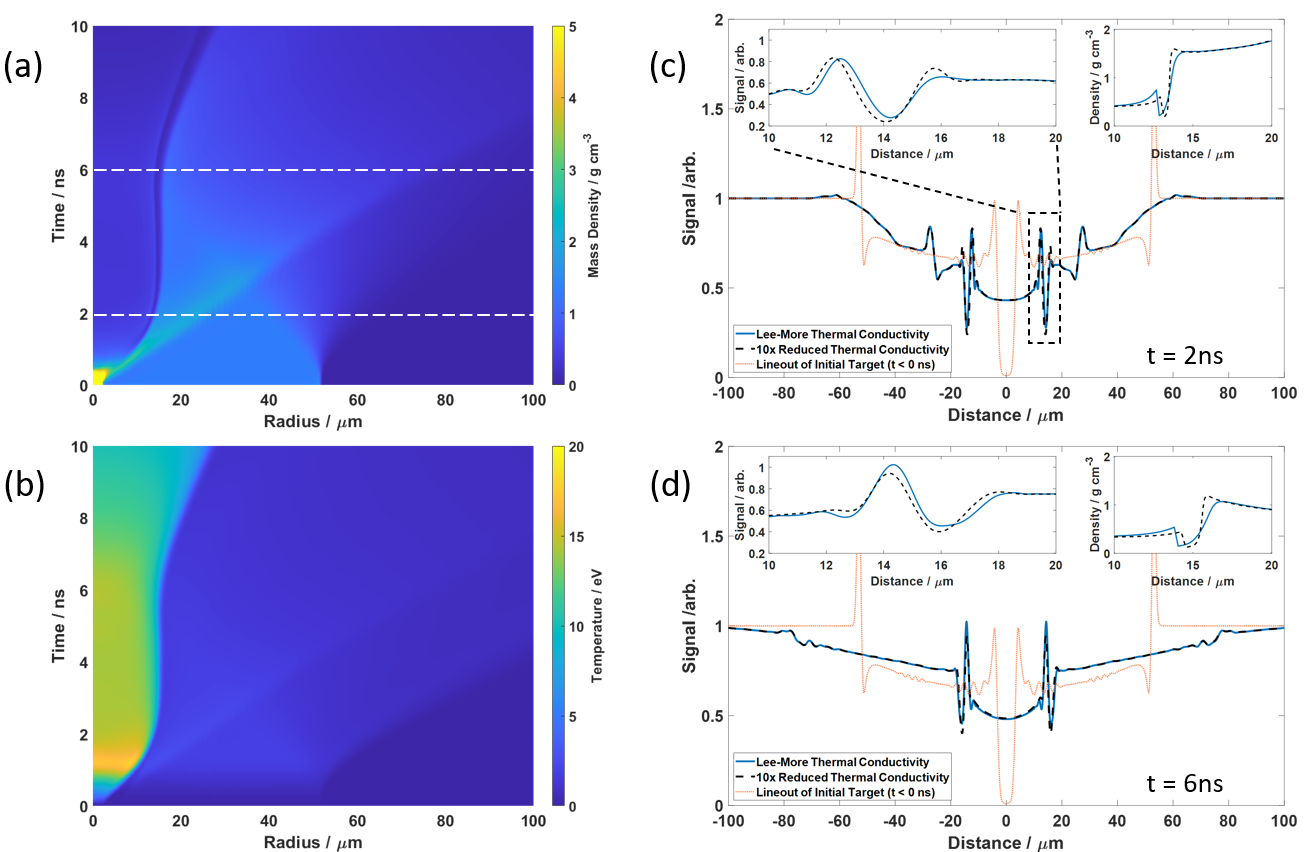}
  \caption{\small{{\bf Target evolution and predicted diffraction patterns.} Hydrodynamic simulations performed in Hydra\cite{Marinak1996} show the evolution of the (a) mass density, and (b) temperature of a 4~$\mu$m W wire coated with 50~$\mu$m CH after irradiation with a $\sim$1~ns 8~keV x-ray pulse. Panels (c) and (d) show the predictions of the diffraction pattern calculated from the target’s complex transmission function at 2 and 6~ns compared to the target prior to heating (t<0~ns). The insets in (c) and (d) detail the shape of the main interface fringes (left) and the radial density profiles (right).} The baseline simulations calculated thermal conductivity using the Lee-More model\cite{LeeMore1984}{}. Simulations that are run with 10x less thermal conductivity show a reduced scale-length at the interface which is captured in the FDR signal. 
  }
 \label{fig:predictions}
\end{figure}

Figures \ref{fig:predictions}a and b show the simulated evolution of the W-CH interface. Prior to heating, the two materials are at their solid densities, and the interface remains a sharp jump from 19.3~g~cm$^{-3}$ to 1.2~g~cm$^{-3}$. As the isochoric heating begins, the shorter attenuation length of W ($\lambda$\textsubscript{mfp} $\sim$3~$\mu$m) causes it to absorb more x-rays per atom than the plastic ($\lambda$\textsubscript{mfp} $\sim$1900~$\mu$m), causing it to  reach higher temperatures. 
We expect an initial temperature in the W of $\sim$15~eV and in the CH of $\sim$5~eV.
The temperature, density, and equation-of-state (EOS) differences between the two materials create a pressure imbalance at the interface; this causes a shock wave to propagate outwards, rapidly reducing the buried W wire's density to $\sim1$g~cm$^{-3}$. Inside the shock front, there is a growing region of near-constant pressure.

After pressure equilibration, the interface's position remains static, but crucially a large temperature differential remains. With no thermal conduction, we would expect a smooth step-like density profile to remain. However, as heat is conducted from the W into the CH, the interface becomes more expansive, and a more complex density profile emerges (see top-right inset in Fig. \ref{fig:predictions}c and d). The change in density is a result of rapid pressure equilibration and inversely mirrors the target's temperature variation. The subsequent evolution of the interface is determined predominantly by thermal conductivity, particularly as the high atomic mass of the W precludes particle diffusion on this timescale. After around 6~ns, the rarefaction wave from the target's exterior reaches the interface, and the target disassembles.

\section{Fresnel Diffractive Radiography}

Current x-ray radiography approaches at large laser facilities use area backlighting methods to generate efficient x-ray sources for imaging by irradiating thin foils\cite{Rygg2014}. By placing a thin slit between the backlighter and target, the x-ray source is reduced in size, allowing for higher resolution via phase-contrast imaging (PCI) or refraction-enhanced radiography (RER) \cite{Ping2011, Dewald2018, Dewald2020}, which is itself based on the work of Pogany {\emph et al.}\cite{Pogany1997, Margaritondo1999}{}. 
In RER, an x-ray source is used to illuminate a target through a slit and the resulting refracted x-rays are recorded on an x-ray framing camera. However, in order to measure the effect of transport properties at the material boundary, the imaging platform must be sensitive to density scale lengths on the order of 1 $\mu$m, particularly for diffusive processes \cite{Haxhimali2014}. This requirement drives the need for an x-ray source size of 1~$\mu$m or smaller. 

To this end, we have developed a method we have termed Fresnel Diffractive Radiography (FDR), an improvement on RER that is capable of resolving these small length scales. By reducing the source size to the order of a micron, we are additionally capable of resolving diffractive effects at sharp density gradients in the target, a notable addition to current RER methods. This is made possible by the size of the slits we have used - novel 1~$\mu$m-wide slits were created specifically for our platform, and are to the authors’ knowledge the smallest slits used at a large laser facility thus far, five times narrower than previous examples \cite{Dewald2020}{}. 

Diffractive effects become important at Fresnel numbers $F = a^2 / (f \lambda) < 1$ \cite{Koch2009}{}, where $1/f=(1/p+1/q)$ with slit-target distance $p = 1.3$~cm, target-image distance $q = 100$~cm, and $\lambda = 2.38~\text{\AA}$ is the wavelength of the V He-$\alpha$ x-rays. At the outer edge of the cylinder, the scale length, $a$, of the plastic-vacuum interface leads to F>1, demonstrating that refraction, rather than diffraction, dominates at this interface. Conversely, the scale length of the tungsten-plastic interface is less than 100~nm, resulting in F$\ll$1 and confirming the importance of including diffractive effects. In fact, in the object plane, the diffractive and refractive fringe widths are comparable ($\sim$~2 $\mu$m). However, in this case, absorption of the W precludes any significant refraction features, leaving absorption and diffraction as the main drivers for image formation. 

Figures \ref{fig:predictions}c and d show predicted diffraction signals at 2~ns and 6~ns, respectively. The baseline simulations calculated thermal conductivity using the Lee-More model\cite{LeeMore1984}. Also shown are simulations employing a 10x decreased conductivity, reducing the density profile's scale-length across the interface. Importantly, this reduction is independent of the choice of equation-of-state. Details of this density variation are captured in the diffraction signal. At 2~ns, this results in a conductivity-sensitive diffraction peak located at around 16~$\mu$m. In contrast, at 6~ns, the effects of thermal conductivity widen the signal. Parameterization of the expected density profile allows for the scale-length of conduction into the two materials to be extracted. 

\section{Experimental Design} \label{expDes}

\begin{figure}
\centering
  \includegraphics[width=1\linewidth]{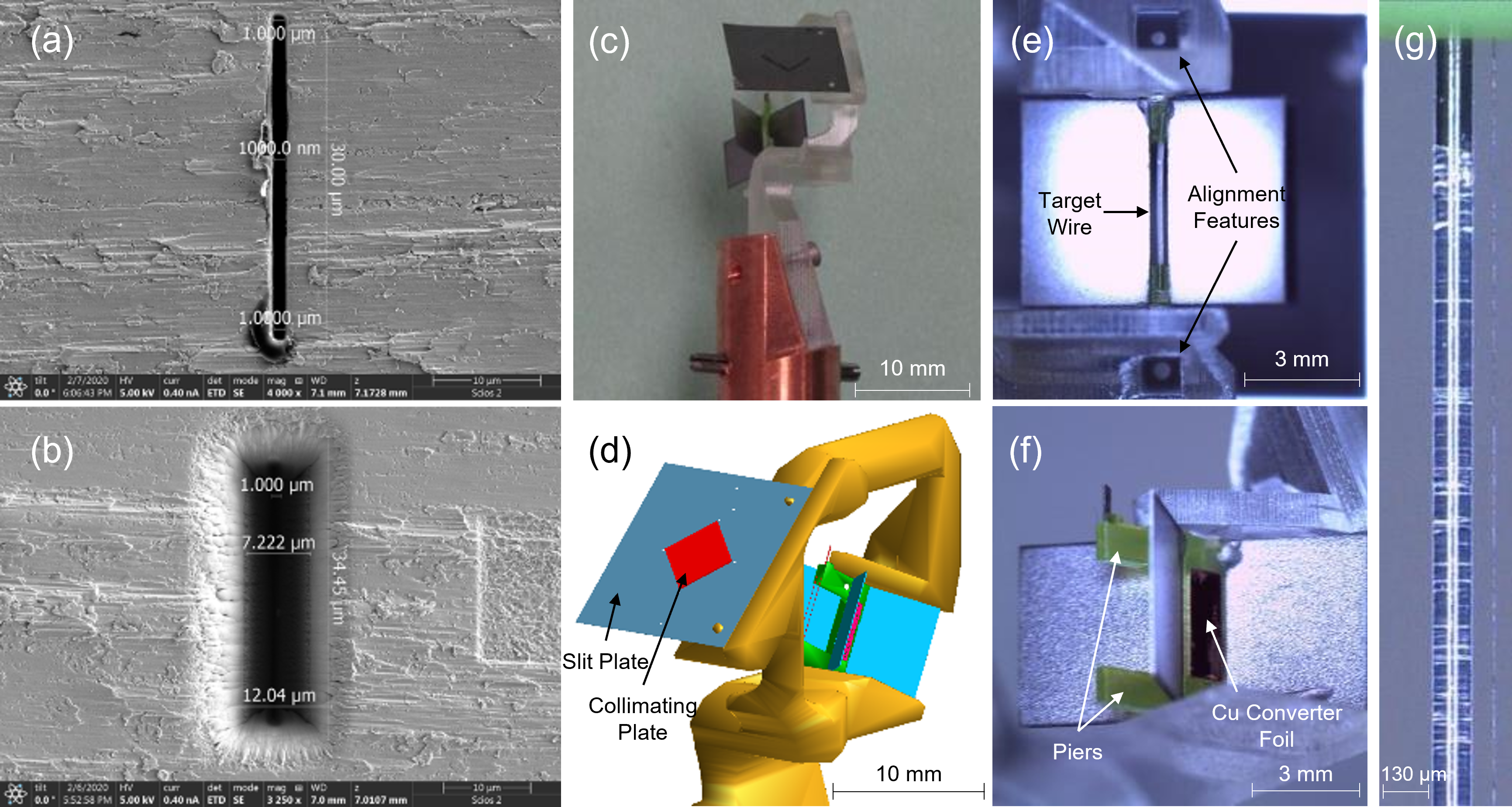}
  \caption{\small{\bf Target montage.} (a-b) SEM images of the tapered 1~$\mu$m wide slit. The slit is cut into a 30~$\mu$m thick, 10~$\times$~10~mm Ta plate using a focused ion beam, tapering from 1~$\mu$m (a) to 10~$\mu$m (b). (c-f) To allow for pre-alignment and characterization of the slit tilt with respect to the wire target, we have developed 3D-printed monolithic structures to hold the slit and target on the same positioner (c). (d) The VisRad model with the various components color-coded, with the slit (dark blue) and physics package (green) being held in the monolithic frame (gold). A collimating plate is placed on the slit plate (red) to help reduce hard x-ray transmission. Multiple alignment holes are cut into the slit plate to align the monolith in the chamber and  perform offline characterization of the wire-slit relative angle, demonstrated in (e). (f) Detail of the physics package, showing the copper converter foil on the side, and the extended piers with fiducial wires. (g) A CH-coated W target wire (OD$\sim$130~$\mu$m) next to an uncoated W wire (OD 4~$\mu$m).}
  \label{fig:targets}
\end{figure}

The success of our experiments crucially hinges on the x-ray source size. This is predominantly solved by the manufacture of our novel 1~$\times$~30~$\mu$m slits\cite{OliverPreprint}. These slits are cut with extremely high precision into 30~$\mu$m thick, 10~$\times$~10~mm square Ta plates using a focused ion beam (FIB) at the University of Nevada, Reno. Given the thickness of the Ta foil, the slit is tapered from 10~$\mu$m (on the side facing the backlighter foil) to 1~$\mu$m (on the side facing the physics target); this serves to help with alignment tolerances  by reducing the impact of any undesired rotation along the axis of the slit. Scanning electron microscope (SEM) images of the slit are shown in Fig. \ref{fig:targets}a and b, which clearly show this tapering. Finally, to reduce slit closure due to x-ray heating, the slits are filled with CH by coating the Ta plate with parylene.

The vertical extent of the slit precludes diffraction along that axis but acts to increase flux. We improve signal-to-noise by integrating vertically along the length of the wire. Due to this, any relative rotation $\theta$ between the slit and the wire increases the effective source size and acts to blur the diffraction pattern. To ensure micron-scale measurements, we require the slit and the wire to be aligned within 1$^\circ$. For small angles, the effective source size $a$ increases by approximately 0.5~$\mu$m for each degree of relative rotation: 
\begin{equation}
a \approx 1~\mathrm{\mu m} + 30~\mathrm{\mu m}  \tan{\theta} .
\label{eq1}
\end{equation}

A number of design choices were made to minimize the impact of any misalignment between the wire and the slit. Primarily, we have developed a single-piece, 3D printed frame ("monolith") with the help of the staff at the Laboratory for Laser Energetics that holds both the slit plate and the physics package, allowing for offline alignment analysis and ensuring a rigid connection between the two, as shown in Fig. \ref{fig:targets}c-f. To get a measure of the relative angle between the slit and the wire, two square holes in the monolith line up with two circular holes in the slit plate, visible in Fig. \ref{fig:targets}e along with the target wire, that are laser cut in line with the vertical axis of the slit. By taking the angle between the center of the circular holes and comparing that to the angle of the wire, we have an approximate measurement of the relative rotation. In the six monolithic targets we have fielded with this feature, listed in Table \ref{table:align}, the worst relative rotation was 0.66$^\circ$ for an effective source size of $\sim$1.35~$\mu$m, and an average relative rotation of 0.36$^\circ$ for an effective source size of $\sim$1.2~$\mu$m. In addition to the wire alignment features, other alignment and mounting holes are laser cut into the Ta slit plate at General Atomics, as seen in the VisRad model in Fig. \ref{fig:targets}d. These additional holes are used in aligning the monolith inside the chamber at OMEGA, as well as a symmetry break to know which direction the slit taper is oriented.

\begin{table}

\centering
\begin{tabular}{|c | c|} 
 \hline
 Monolithic Frame & Measured Relative Tilt ($^\circ$) \\
 \hline
 TND-OMDIFF-76 & 0.47 $\pm$ 0.01  \\ 
 \hline
 TND-OMDIFF-77 & 0.25 $\pm$ 0.01  \\
 \hline
 TND-OMDIFF-78 & 0.56 $\pm$ 0.01  \\
 \hline
 TND-OMDIFF-79 & 0.08 $\pm$ 0.01 \\
 \hline
 TND-OMDIFF-80 & 0.15 $\pm$ 0.01 \\
 \hline
 TND-OMDIFF-84 & 0.66 $\pm$ 0.01 \\
 \hline  
 Mean and Std. Dev. & 0.36 $\pm$ 0.21 \\
 \hline
\end{tabular}
\caption{\small{{\bf Monolithic Frames and the measured tilt between the slit and target wire.} Six monolithic targets were used with alignment features to measure the relative tilt between the slit and the target wire. The measured tilts are all well within 1$^\circ$, minimizing source broadening due to tilt.}}
\label{table:align}
\end{table}

The physics package is assembled separately and is inserted into the monolith afterwards. This 3D printed piece holds all of the components for the isochoric heating together, including the Cu x-ray converter foils and the target wires. Two piers extend outwards towards the slit plate and hold uncoated W wires as imaging fiducials, most easily seen in Fig. \ref{fig:targets}f. Arrayed outwards from the physics package and flanking the Cu foils are four 4~mm long~$\times$~5~mm tall 50~$\mu$m thick Ta foil "shields" that serve to reduce x-ray emission from the Cu foils from reaching the detector or heating the fiducial wires.
The final component of our physics package is the target wire. These are nominal 4~$\mu$m W wires sourced from Goodfellow (diameter tolerance $\pm$ 10$\%$) that have been coated with $\text{C}_{\text{8}}\text{H}_{\text{4}}\text{F}_{\text{4}}$ via physical vapor deposition at the Central Laser Facility in the UK. A coated wire can be seen in the physics package in Fig. \ref{fig:targets}e, while a detailed example is shown in Fig. \ref{fig:targets}g; an uncoated W wire is shown immediately to the right for comparison. The total diameter for the coated wire is $\sim$130~$\mu$m. 

The dimensions of the monolith are designed to allow for a fixed distance between the slit and the target wire of 13~mm. With the x-ray framing camera (XRFC) placed 1000~mm further behind the target wire, we have a magnification M = 1000 / 13 $\approx$ 77. For a detector area of approximately 34~x~34~mm we image $\sim$450~x~450~$\mu$m at the target plane, allowing us to see both the W-CH and CH-vacuum interfaces, and have a significant vertical length of the wire to integrate across. 

\subsection{Facility Configuration}
The current iteration of the monolith frame is specifically designed to interface with the OMEGA planar cryogenic carts. While none of the targets themselves are cryogenic, the ability to use the planar cryo-carts as a target positioner allowed for the additional bulk of the monolith over the use of stalk-mounted targets, while maintaining the desired imaging axis. The monolith shown in Fig. \ref{fig:targets}c features the copper base at the bottom of the 3D printed frame that connects the target to the planar cryo-cart.

These experiments use 26 of OMEGA's 60 laser beams\cite{OMEGAFacility} at 3~$\omega$, 10 of which are used for driving the backlighter, and the remaining 16 are used to heat the target wire into WDM conditions. For all beams, we use a 1~ns square pulse, and delay the backlighter beams from the start of the drive beams by up to 8~ns to probe the wire target as it evolves in time. The 10 backlighter beams are tuned to 450~J/beam and focused to the same spot on a V backlighter foil. Due to the small size of the slit, we worked diligently to ensure our backlighter had sufficient energy to provide us with enough photons to perform our experiments. If we assume 4.5~kJ of energy from the backlighter beams with 0.35\% conversion efficiency into 5.2~keV V x-rays\cite{HuntingtonRSI}, we can expect 10$^{16}$ total photons emitted into 4$\pi$. A slit with dimensions of 1~$\mu$m~$\times$~30~$\mu$m located 5~mm away transmits $\sim$10$^{11}$ photons, resulting in 200 photons incident on a 1~$\mu$m~$\times$~1~$\mu$m element of the wire. Accounting for 35\% transmission through target and filters, 5\% micro-channel plate detective quantum efficiency, gating over 0.24~ns, and averaging over the 450~$\mu$m long wire, that is $\sim$1500 photons/1~$\mu$m resolution element and a SNR$\approx$50. To additionally isolate 5.2~keV V x-ray emission, a Ross-Pair filter consisting of 7~$\mu$m V and 10~$\mu$m Ti foils is used to filter out other photon energies. 

The 16 drive beams are divided into 8 beams on each side of the target, and pointed in groups of 4 to create a larger overlapping beam spot on the Cu x-ray converter foils. The irradiated area on each of these converter foils is approximately 0.6~mm$^2$, much larger than the corresponding area of the target wire at 0.13~mm$^2$. With each beam tuned to a maximum of 450~J/beam, this results in intensities on the order of 10$^{15}$~W~cm$^{-2}$. The relative size difference between the drive spots and the target wire diameter creates an x-ray bath that can heat the target near uniformally. The aforementioned Ta shielding is used to block any x-ray emission from the drive spot from reaching the detector.


\begin{figure}
\centering
  \includegraphics[width=1\linewidth]{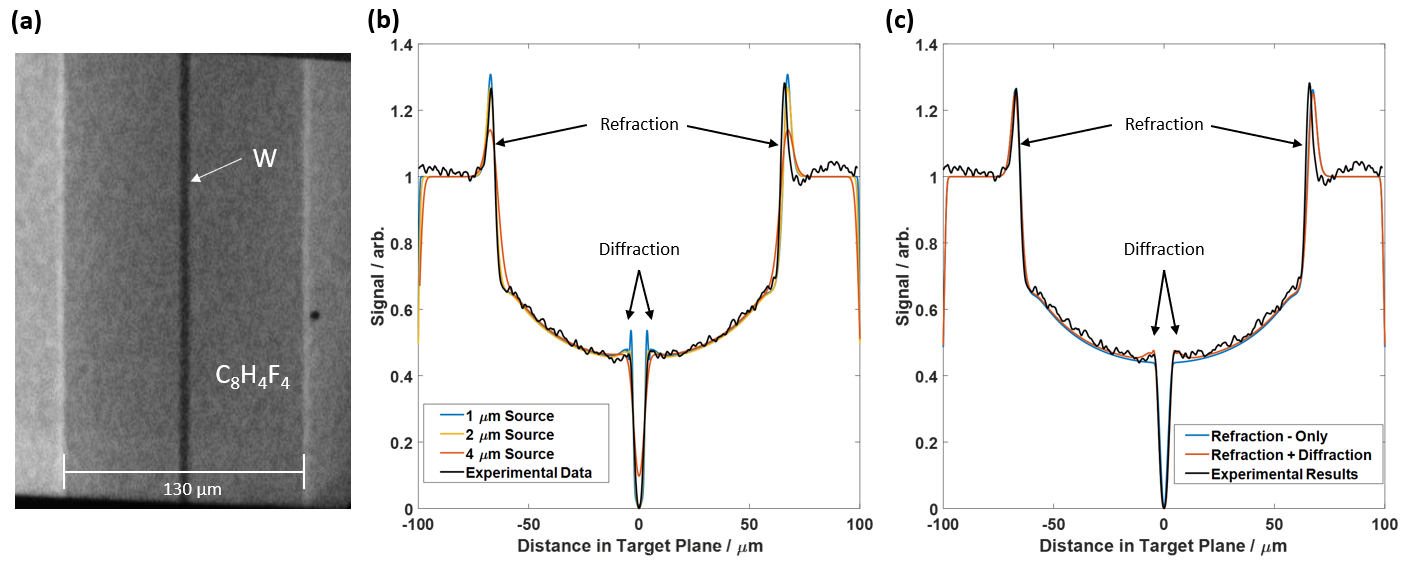}
  \caption{\small{\bf Proof of principle measurement.} (a) Radiograph showing a 4~$\mu$m W wire inside a 130~$\mu$m diameter $\text{C}_{\text{8}}\text{H}_{\text{4}}\text{F}_{\text{4}}$ wire before isochoric heating. The W wire is visible in the center of the image, as is the CH coating. (b) A lineout of the integrated length of the coated wire is compared to several Gaussian sources with specified sizes. As the source size approaches the size of the W wire, refraction and diffraction features become less sharp, and we begin to see signal through the W wire. (c) The same lineout of the data is compared to best-fit predictions using a full refraction-diffraction treatment \cite{Pogany1997} and a simplified refraction-only approximation \cite{Koch2013}, and shows a clear distinction at the W-CH interface necessitating the full diffraction calculation.}  
  \label{fig:prelim}
\end{figure}

\section{Results}
The use of monolithic frames for our FDR imaging platform has allowed for improved characterization of our targets by facilitating offline alignment analysis.  Results from an undriven shot (100082, using Frame TND-OMDIFF-78) are shown in Fig. \ref{fig:prelim}a, with the CH-coated W wire and its different interfaces easily visible in the raw image. In Fig. \ref{fig:prelim}b and \ref{fig:prelim}c, a lineout integrated vertically across Fig. \ref{fig:prelim}a is is compared to various simulations of a concentric, symmetrically coated wire. These simulations are generated using the known and measured parameters of the experiment, including the measured angle from Table \ref{table:align} for TND-OMDIFF-78, a measured $\text{C}_{\text{8}}\text{H}_{\text{4}}\text{F}_{\text{4}}$ density of $\sim$1.85~g~cm$^{-3}$, a measured total diameter of $\sim$130~$\mu$m, and a W wire diameter of 4.2~$\mu$m consistent with the tolerances provided by Goodfellow. 

Fig. \ref{fig:prelim}b compares the experimental results to a series of perfect Gaussian x-ray sources with a FWHM source size of 1, 2, and 4~$\mu$m. We can see that for source sizes smaller than the W wire, we have strong diffraction and refraction features, and a fully opaque W wire. For source sizes on order of the width of the W wire, we lose the diffraction features at the W-CH interface, and predict significant signal through the W. These results indicate that we have a source size of approximately 2~$\mu$m. However, using the measured relative rotation and Eq. \ref{eq1}, we predict a source size of approximately 1.3~$\mu$m. We believe that the source is additionally broadened by x-ray transmission through the narrow end of the slit tapering. In future target designs, we will implement more sophisticated tapering profiles to minimize this source broadening effect.

Fig. \ref{fig:prelim}c compares the experimental results to two different  calculations, one that includes a full solution to the Fresnel-Kirchhoff Diffraction Integral \cite{Pogany1997}{} (labeled Refraction+Diffraction) and another that uses a simplified refraction-only approximation \cite{Koch2013}{}. As described in Section 3, the outside edge of the cylinder is well described by both calculations, but the inner W-CH interface has a clear distinction between the two that demonstrates the existence of diffraction features and the need for the more complete calculation.

\section{Summary and Outlook}
Our preliminary analysis has shown that FDR is capable of achieving high spatial resolutions and resolving diffraction features resulting from the edge of narrow, high absorption materials. The spatial resolution is not as high as expected, in large part due the transmission of X-rays through the tapered parts of the 1~$\mu$m Ta slits. For future experiments, we will be adjusting the design of the slit tapering profile to account for this. For instance, by reducing the tapering width from 10~$\mu$m to 5~$\mu$m or less, we expect to achieve 1~$\mu$m spatial resolution, albeit at the cost of alignment tolerances. 

Analysis for the driven shots is ongoing, requiring more in-depth simulation to allow forward-fitting and extraction of physical properties like the thermal conductivity of the materials. The favorable comparison between our simulated model and the undriven data gives confidence in our ability to accurately predict the evolution of our targets after isochoric heating.

The undriven shot provides a good indicator of the signal-to-noise ratio we can expect using this platform. 
We find the experimentally measured noise in the FDR signal is <~2\%. Comparatively, the difference in the predicted diffraction pattern signal for the two thermal conductivity values plotted in Fig. 2 is approximately 13\%. Thus, we estimate that we can determine variation in the thermal conductivity within approximately 2.5x. These results will help benchmark theoretical models and supplement existing observations such as those given by Sugimoto et al.\cite{Sugimoto2017}, which vary by over an order of magnitude.

Furthermore, based on the preliminary success of the monolithic platform, we will be continuing to refine the design for upcoming experiments at the OMEGA Laser Facility. The ability to characterize the relative angle between the slit and target wire in advance of the shot helps to reduce uncertainty in the analysis, particularly in understanding the effective source size based on the data. 



\begin{backmatter}
\bmsection{Funding} The work of Y.P., L.D., O.L., M.S., A.K., and T.D. was performed under the auspices of the U.S. Department of Energy by Lawrence Livermore National Laboratory under Contract DE-AC52-07NA27344 and supported by Laboratory Directed Research and Development (LDRD) Grant No. 21-ERD-029.

This material is based upon work supported by the National Science Foundation under Grant No. PHY-2045718

\bmsection{Acknowledgments} The authors would like to thank Roger Janezic, Jack Armstrong, James Tellinghuisen, and the entire OMEGA 60 staff for all of their assistance in developing and fielding our monolithic targets inside the chamber. The authors would also like to thank Zachary Karmiol at the University of Nevada, Reno for machining and characterizing all of the slits we have used in our experiments. 

\bmsection{Disclosures} The authors declare no conflicts of interest.

\bmsection{Data availability} Data underlying the results presented in this paper are not publicly available at this time but may be obtained from the authors upon reasonable request.
\end{backmatter}

\newpage
{
}


\begin{thebibliography}{10}

\bibitem{Guillot1999}
T. Guillot. "A comparison of the interiors of Jupiter and Saturn." Planetary and Space Science {\bf 47}.10-11 (1999): 1183-1200.

\bibitem{Paquette1986}
C. Paquette, C. Pelletier, G. Fontaine, and G. Michaud. "Diffusion Coefficients for Stellar Plasmas." Astrophys. J. Suppl. Ser. {\bf 61} (1986): 177.

\bibitem{Atzeni2004}
S. Atzeni and J. Meyer-ter-Vehn. "The Physics of Inertial Fusion: Beam Plasma Interaction, Hydrodynamics, Hot Dense Matter." Oxford University Press on Demand {\bf 125} (2004).

\bibitem{Hurricane2014}
O. Hurricane, D. Callahan, D. Casey, \emph{et al.} "Fuel gain exceeding unity in an inertially confined fusion implosion." Nature {\bf 506} (2014): 343–348. 


\bibitem{Hu2016}
S. X. Hu, L. A. Collins, V. N. Goncharov, J. D. Kress, R. L. McCrory, and S. Skupsky. "First-principles investigations on ionization and thermal conductivity of polystyrene for inertial confinement fusion applications." Physics of Plasmas {\bf 23}.4 (2016): 042704. 

\bibitem{Grabowski2020}
P.E. Grabowski, S. B. Hansen, M. S. Murillo, \emph{et al.} "Review of the first charged-particle transport coefficient comparison workshop." High Energy Density Physics, {\bf 37} (2020): 100905

\bibitem{Graziani2014}
F. Graziani, M. P. Desjarlais, R. Redmer, and S. B. Trickey, eds. Frontiers and challenges in warm dense matter. Vol. 96. Springer Science \& Business, (2014).

\bibitem{McKelvey2017}
A. McKelvey, G. E. Kemp, P. A. Sterne, \emph{et al.} "Thermal conductivity measurements of proton-heated warm dense aluminum." Scientific Reports {\bf 7}.1 (2017): 1-10.

\bibitem{Ping2019}
Y. Ping, H. D. Whitley, A. McKelvey, \emph{et al.} "Heat-release equation of state and thermal conductivity of warm dense carbon by proton differential heating." Physical Review E {\bf 100}.4 (2019): 043204.

\bibitem{Sugimoto2017}
S. Sugimoto, A. Watabe, Y. Sugimoto, \emph{et al.} "Observation of the thermal conductivity of warm dense tungsten plasma generated by a pulsed-power discharge using laser-induced fluorescence." Physics of Plasmas {\bf 24}.7 (2017): 072703.



\bibitem{OMEGAFacility}
T. R. Boehly, R. S. Craxton, T. H. Hinterman, \emph{et al.} “The Upgrade to the OMEGA Laser System,” Proc. SPIE {\bf 1627} (1992): 236-245.

\bibitem{Rygg2014}
J. R. Rygg, O. S. Jones, J. E. Field, \emph{et al.} "2D X-Ray Radiography of Imploding Capsules at the National Ignition Facility" Phys. Rev. Lett. {\bf 112} (2014): 195001.

\bibitem{Ping2011}
Y. Ping, O. L. Landen, D. G. Hicks, \emph{et al.} "Refraction-enhanced x-ray radiography for density profile measurements at CH/Be interface." Journal of Instrumentation, {\bf 6} (2011): P09004

\bibitem{Dewald2018}
E. L. Dewald, O. L. Landen, L. Masse, \emph{et al.} "X-ray streaked refraction enhanced radiography for inferring inflight density gradients in ICF capsule implosions." Review of Scientific Instruments, {\bf 89} (2018): 10G108

\bibitem{Dewald2020}
E. L. Dewald, O. L. landen, D. Ho, \emph{et al.} "Direct observation of density gradients in ICF capsule implosions via streak Refraction Enhanced Radiography (RER)." High Energy Density Physics, {\bf 36} (2020): 100795

\bibitem{Pogany1997}
A. Pogany, D. Gao, and S. W. Wilkins. "Contrast and resolution in imaging with a microfocus x-ray source." Review of Scientific Instruments {\bf 68}.7 (1997): 2774-2782. 

\bibitem{Margaritondo1999}
G. Margaritondo and G. Tromba. "Coherence-based edge diffraction sharpening of x-ray images: a simple model." Journal of Applied Physics {\bf 85}.7 (1999): 3406-3408.

\bibitem{Haxhimali2014}
T. Haxhimali, and R. Rudd. "Diffusivity of mixtures in warm dense matter regime" Frontiers and Challenges in Warm Dense Matter. Springer, Cham, 235-263 (2014).

\bibitem{Marinak1996}
M. M. Marinak, R. E. Tipton, O. L. Landen, \emph{et al.} "Three‐dimensional simulations of Nova high growth factor capsule implosion experiments" Physics of Plasmas {\bf 3}, (1996): 2070

\bibitem{LeeMore1984} 
Y. T. Lee and R. M. More. "An electron conductivity model for dense plasmas." The Physics of Fluids {\bf 27} (1984): 1273

\bibitem{Koch2009}
J. A. Koch, O. L. Landen, B. J. Kozioziemski, \emph{et al.}"Refraction-enhanced x-ray radiography for interial confinement fusion and laser-produced plasma applications." Journal of Applied Physics {\bf 105} (2009): 113112

\bibitem{OliverPreprint}
M. Oliver, C. H. Allen, L. Divol, \emph{et al.} "Diffraction Enhanced Radiography with Laser-Produced X-ray Sources." Preprint (2022).

\bibitem{HuntingtonRSI}
C. M. Huntington, C. M. Krauland, C. C. Kuranz, \emph{et al.} "Development of a short duration backlit pinhole for radiography on the National Ignition Facility." Rev. Sci. Instrum. {\bf 81} (2010): 10E536.

\bibitem{Koch2013}
J. A. Koch, O. L. Landen, L. J. Suter, and L. P. Masse. "Simple solution to the Fresnel-Kirchoff diffraction integral for application to refraction-enhanced radiography." J Opt Soc Am A {\bf 30}.7 (2013): 1460-1463.



\end{thebibliography}
\end{document}